\documentclass[12pt]{iopart}
\usepackage{graphicx}
\usepackage{latexsym}
\usepackage{url}

\usepackage{iopams}
\def\aap{A\&A\,  }
\def\aj{AJ  }
\def\apj{ApJ\,  }
\def\apjl{ApJ\,  }
\def\apjs{ApJS  }

\def\nat{Nature\,  }

\def\h0units{\mathrm{km\,s^{-1}\,Mpc^{-1}}}
\def\cunits{\mathrm{km\,s^{-1}}}

\newcommand{\om}{\Omega_{\rm M}}

\newcommand{\dl}{d_{\rm{L}}}
\begin{document}
\title
{
An analytical solution in the complex plane
for the luminosity
distance in flat cosmology
}
\vskip  1cm
\author     {Lorenzo Zaninetti}
\address    {Physics Department  ,
 via P.Giuria 1,\\ I-10125 Turin,Italy }
\ead {zaninetti@ph.unito.it}

\begin {abstract}
We present an  analytical solution for the luminosity distance
in spatially flat cosmology with pressureless matter and
the cosmological constant.
The complex analytical solution is made of  a real part
and a negligible imaginary part.
The real part of the luminosity distance allows finding
the two parameters $H_0$ and $\om$.
A simple expression for the distance modulus for SNs of type Ia 
is reported in the framework of the minimax approximation.
\end{abstract} 
\vspace{2pc}
\maketitle  

\section{Introduction}

The luminosity distance in flat cosmology has been recently
investigated using different approaches.
A fitting formula which has a maximum relative error of 
$4\%$ in the case of common cosmological parameters has been
introduced by \cite{Pen1999}.
An approximate solution in terms 
of Pad\'e  approximants
has been presented by \cite{Adachi2012}.
The integral of the luminosity distance has been
found in terms of elliptical integrals of the first kind
by \cite{Meszaros2013}.

\section{Flat cosmology}

Following Eq. (2.1) in \cite{Adachi2012},
the luminosity distance $\dl$ is
\begin{equation}
  \dl(z;c,H_0,\om) = \frac{c}{H_0} (1+z) \int_{\frac{1}{1+z}}^1
  \frac{da}{\sqrt{\om a + (1-\om) a^4}} \quad ,
  \label{lumdistflat}
\end{equation}
where $H_0$
is the Hubble constant expressed in     $\h0units$,
$c$ is the speed  of light expressed in $\cunits$,
$z$ is the redshift,
$a$ is the scale-factor,
and  $\om$ is
\begin{equation}
\om = \frac{8\pi\,G\,\rho_0}{3\,H_0^2}
\quad ,
\end{equation}
where $G$ is the Newtonian gravitational constant and
$\rho_0$ is the mass density at the present time.
We now introduce the indefinite integral
\begin{equation}
\Phi(a) = \int
  \frac{da}{\sqrt{\om a + (1-\om) a^4}}
\quad .
\end{equation}
The solution is in terms of
$\mathop{F\/}$, the Legendre integral or incomplete
elliptic integral of the first kind
\begin{equation}
\Phi(a) = \frac
{
-4\, \mathop{F\/} \left( b_{{1}},b_{{2}} \right) b_{{3}}b_{{4}}b_{{6}}b_{{1}}b_{{5
}}
}
{
b_{{7}}b_{{8}}\sqrt {b_{{9}}}b_{{10}}
}
\quad ,
\end{equation}
where the incomplete elliptic integral of the first kind is
\begin{equation}
\mathop{F\/}\nolimits\!\left(x,k\right)=\int_{0}^{\mathop{x}}\frac{dt}{\sqrt{1-t^{2}}\sqrt{1-k^{2}t^{2}}}
\quad ,
\end{equation}
see formula (19.2.4) in \cite{NIST2010},
and
\begin{equation}
b_1= \sqrt {-{\frac {a \left( {\it \om}-1 \right)  \left( i\sqrt {3}+3
 \right) }{ \left( -{\it \om}\,a+\sqrt [3]{{\it \om}\, \left( {\it \om}-1
 \right) ^{2}}+a \right)  \left( i\sqrt {3}+1 \right) }}}
\quad ,
\nonumber
\end{equation}
\begin{equation}
b_2 = \sqrt {{\frac { \left( i\sqrt {3}+1 \right)  \left( i\sqrt {3}-3
 \right) }{ \left( i\sqrt {3}+3 \right)  \left( i\sqrt {3}-1 \right) }
}}
\quad ,
\nonumber\end{equation} 	
\begin{equation}
b_3 =
\sqrt {{\frac {i\sqrt {3}\sqrt [3]{{\it \om}\, \left( {\it \om}-1
 \right) ^{2}}+2\,{\it \om}\,a+\sqrt [3]{{\it \om}\, \left( {\it \om}-1
 \right) ^{2}}-2\,a}{ \left( -{\it \om}\,a+\sqrt [3]{{\it \om}\, \left(
{\it \om}-1 \right) ^{2}}+a \right)  \left( i\sqrt {3}+1 \right) }}}
\quad ,
\nonumber\end{equation}
\begin{equation}
b_4 =
\sqrt {{\frac {-i\sqrt {3}\sqrt [3]{{\it \om}\, \left( {\it \om}-1
 \right) ^{2}}+2\,{\it \om}\,a+\sqrt [3]{{\it \om}\, \left( {\it \om}-1
 \right) ^{2}}-2\,a}{ \left( -\sqrt [3]{{\it \om}\, \left( {\it \om}-1
 \right) ^{2}}+a \left( {\it \om}-1 \right)  \right)  \left( i\sqrt {3}
-1 \right) }}}
\quad ,
\nonumber\end{equation}
\begin{equation}
b_5=
i\sqrt {3}+1
\quad ,
\nonumber\end{equation} 	
\begin{equation}
b_6 =
 \left( -{\it \om}\,a+\sqrt [3]{{\it \om}\, \left( {\it \om}-1 \right) ^{
2}}+a \right) ^{2}
\quad ,
\nonumber\end{equation}
\begin{equation}
b_7 =
\sqrt [3]{{\it \om}\, \left( {\it \om}-1 \right) ^{2}}
\quad ,
\nonumber\end{equation}
\begin{equation}
b_8 =
i\sqrt {3}+3
\quad ,
\nonumber\end{equation} 	
\begin{equation}
b_9 =
 \left( -4\,{a}^{4}+4\,a \right) {\it \om}+4\,{a}^{4}
\quad ,
\nonumber\end{equation}
\begin{equation}
b_{10} =
{\it \om}-1
\quad ,
\nonumber\end{equation}
with $i^2=-1$.
The incomplete elliptic integral
$\mathop{F\/}\nolimits\!\left(x,k\right)$
of complex arguments is  evaluated according to
Eq. (17.4.11) in 
\cite{Abramowitz1965} or Section 19.7 (ii)
in \cite{NIST2010}.
The luminosity distance is
\begin{equation}
  \dl(z; c,H_0,\om) =\Re \bigg ( \frac{c}{H_0} (1+z) \big( \Phi(1) -\Phi(\frac{1}{1+z})
\big) \bigg) \quad ,
\end{equation}
where $\Re$ means the  real part.

The distance modulus is
\begin{equation}
(m-M) =25 +5 \log_{10}\bigg ( \dl(z; c,H_0,\om) \bigg)
\quad .
\label{distmod_elliptic}
\end{equation}
An approximation can be found  when
the argument of the integral (\ref{lumdistflat})
is expanded about a=1 in a Taylor series of order 10.
The resulting Taylor approximation of 
order 10 to the luminosity distance, $\dl(z; c,H_0,\om)_{10} $, is
\begin{eqnarray}
  \dl(z; c, H_0,\om)_{10} = \nonumber  \\
  {\frac {c   ( 1+z   ) }{H_{{0}}}}  
  \bigg ( \frac{1}{2}\,   ( \frac{3}{2}\,{\it
\om}-2   )    ( 1-   ( 1+z   ) ^{-2}   ) +3 -3\,
   ( 1+z   ) ^{-1}\nonumber \\ 
-\frac{3}{2}\,{\it \om}\,   ( 1-   ( 1+z
   ) ^{-1}   )  \bigg  )  + \ldots
   \end{eqnarray}
where we have reported the first few terms of the series.
The    goodness of the Taylor  approximation is evaluated
through the percentage error, $\delta$,
which is
\begin{equation}
\delta = \frac{\big | \dl(z; c,H_0,\om) - \dl(z; c,H_0,\om)_{10} \big |}
{\dl(z; c,H_0,\om ) } \times 100
\quad .
\end{equation}
As an example when
$H_0=70 \h0units$,
$\om=0.3$,
$c=299792.458 \cunits$
and $z=4$,  we obtain
$\delta=0.61 \%$.
As an example with the above  parameters,
$\dl$  has  its angle  in the complex plane, 
$\theta$, very small:
$\theta \approx  10^{-11}$, which means that the solution is real
for practical purposes.
In the last years the Hubble Space Telescope (HST) has allowed the determination
of the cosmological parameters 
through the modulus of the distance for SNs of type Ia,
see \cite{Perlmutter1998,Garnavich1998,Riess1998,Knop2003,Riess2007}.
At the moment of writing the two unknown parameters, $H_0$ and  $\om$,
can be derived from two catalogs for  the distance modulus
of SNs  of type Ia: 580 SNe  in  the Union 2.1 compilation, 
see \cite{Suzuki2012} 
with data at \url{http://supernova.lbl.gov/Union/} ,
and   740 SNe  in
the joint light-curve analysis (JLA), 
see \cite{Betoule2014}
with data at \url{http://supernovae.in2p3.fr/sdss_snls_jla/ReadMe.html}.
This
kind of analysis is not new and has been used,
for example,  by \cite{Oliveira2016}.

\begin{figure}
\begin{center}
\includegraphics[width=10cm]{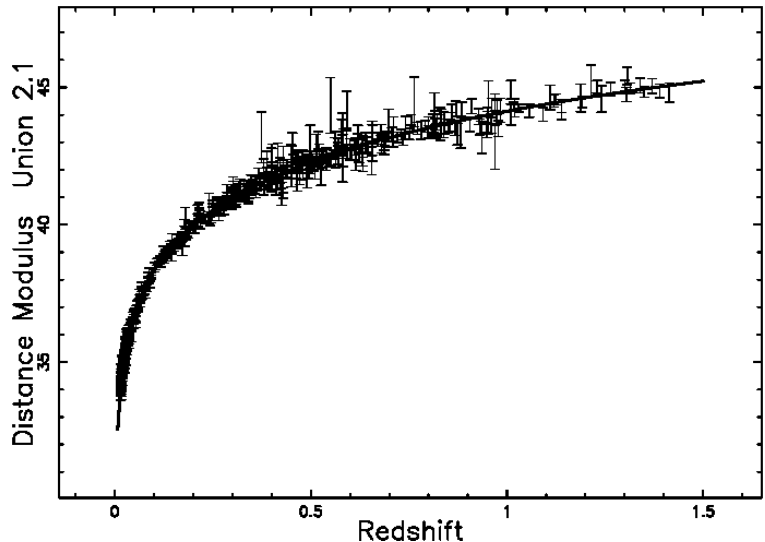}
\end{center}
\caption{
Hubble diagram for the  Union 2.1  compilation.
The solid line represents the best fit
for the exact distance modulus  in flat cosmology 
as represented by Eq. (\ref{distmod_elliptic}),
parameters as in first line of Table \ref{chi2valueflat}.
}
\label{distmoddlflat}
\end{figure}

\begin{figure}
\begin{center}
\includegraphics[width=10cm]{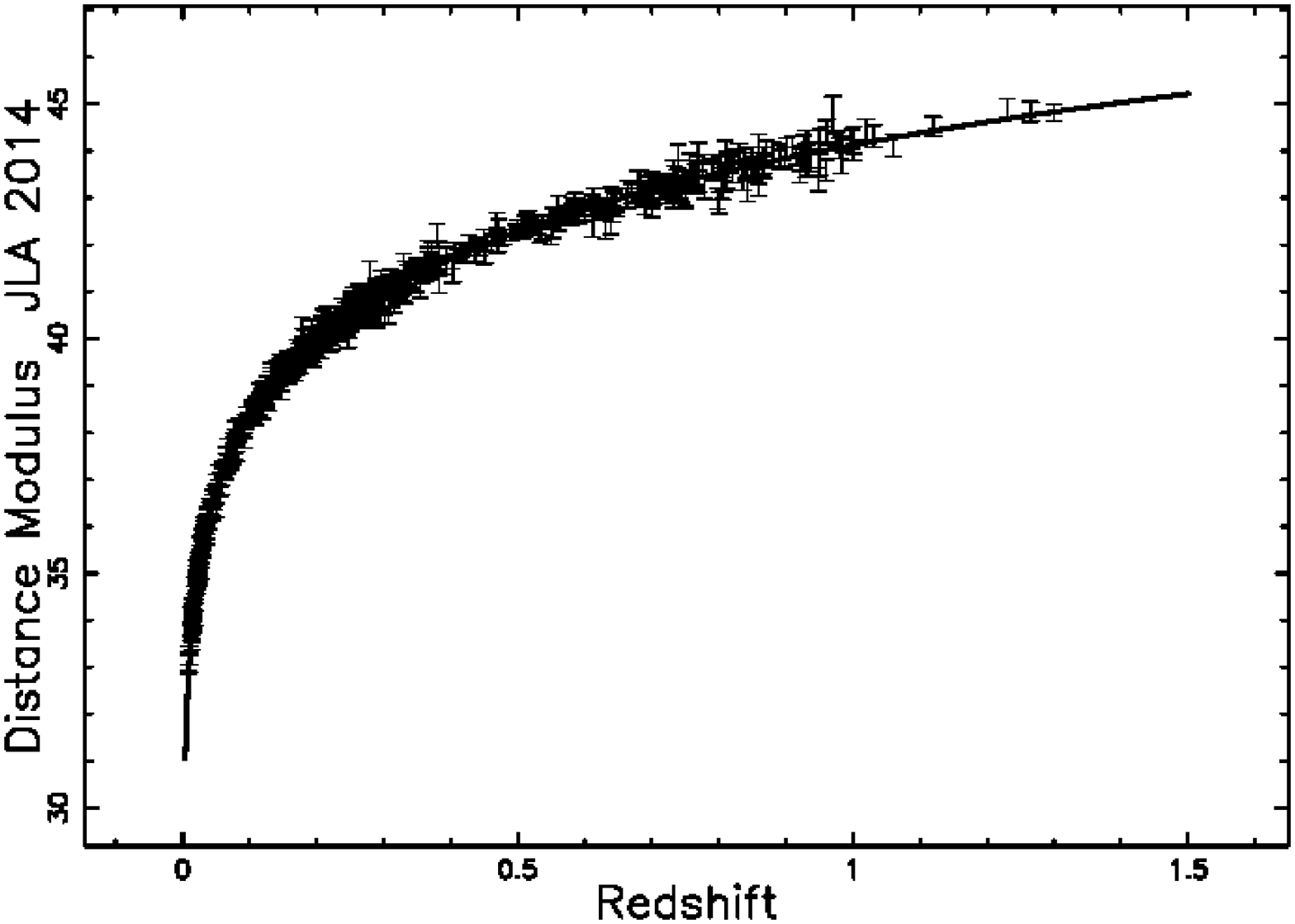}
\end{center}
\caption{
Hubble diagram for the  JLA   compilation.
The solid line represents the best fit
for the exact distance modulus  in flat cosmology as represented by
Eq. (\ref{distmod_elliptic}),
parameters as in second line of  Table \ref{chi2valueflat}.
}
\label{distmoddlflat_jla}
\end{figure}

The best fit for the distance modulus
of SNs is obtained adopting the
Levenberg--Marquardt  method
(subroutine MRQMIN in \cite{press}).
The statistical parameters here adopted are
the merit function or chi-square, $\chi^2$,
the reduced chi-square, $\chi_{red}^2$ and
the maximum probability of obtaining a better fitting, $Q$,
see  Section 2.3 in \cite{Zaninetti2016a} for more details.
Table \ref{chi2valueflat} reports $H_0$ and $\om$
for the two catalogs of SNs and
Figures \ref{distmoddlflat} and \ref{distmoddlflat_jla}
display the best  fits.

\begin{table}[ht!]
\caption
{
Numerical values of
$\chi^2$,
$\chi_{red}^2$
and 
$Q$ where 
$k$ stands for the number of parameters.
}
\label{chi2valueflat}
\begin{center}
\resizebox{12cm}{!}
{
\begin{tabular}{|c|c|c|c|c|c|c|}
\hline
compilation   &  SNs& $k$    &   parameters    & $\chi2$& $\chi_{red}^2$
&
$Q$      \\
\hline
Union~2.1 & 580 &  2
& $H_0=70\pm 0.34$  ; $\om=0.277\pm0.019$
& 562.22 &  0.972  & 0.673 \\
\hline
JLA  & 740 &  2
& $H_0=69.83\pm 0.31$  ; $\om=0.287 \pm0.018$
& 627.82  &  0.85  & 0.998   \\
\hline
\end{tabular}
}
\end{center}
\end{table}

The  Taylor approximation of order 10 to the distance modulus,
$\dl(z; c,H_0,\om)_{10} $, is
\begin{equation}
(m-M)_{10}  =25 +5 \log_{10}\bigg ( \dl(z; c,H_0,\om)_{10} \bigg)
\quad .
\label{distmod_elliptic_taylor}
\end{equation}
The above equation
takes  a simple expression when the
minimax rational approximation is used,
see \cite{Remez1934,Remez1957,NIST2010};
here we have used a polynomial of degree 3 for the numerator
and degree  2 for the denominator.
 the parameters of Table  \ref{chi2valueflat}
for the Union 2.1 compilation
over the range in $z\in [0,4]$,  we obtain
the following minimax approximation
\begin{eqnarray}
(m-M)_{3,2,10} = 
\frac
{
0.413991+ 6.080622\,z+ 5.501967\,{z}^{2}+ 0.029254\,{z
}^{3}
}
{
0.012154 + 0.148352 \,z+ 0.112017 \,{z}^{2}
}
\label{distmodminimaxunion21}
\\
 \quad Union~2.1~compilation
\quad ,
\nonumber  
\end{eqnarray}
the maximum error being 0.002956.

\section{Conclusions}

We have presented an analytical approximation
for the luminosity distance in terms of elliptical
integrals with complex argument.
The fit of the distance modulus of SNs of type Ia allows
finding the pair $H_0$ and $\om$
for the Union 2.1 and JLA compilations.
A simple expression for the distance modulus 
relative to the Union 2.1 compilation is given 
through the minimax approximation 
applied to a Taylor expansion of the luminosity distance
of order 10.

{\bf References}

\providecommand{\newblock}{}

\providecommand{\newblock}{}
\providecommand{\newblock}{}

\end{document}